# People Lie, Actions Don't! Modeling Infodemic Proliferation Predictors among Social Media Users

Chahat Raj[1], Priyanka Meel[2]
[1]chahatraj58@gmail.com, [2]priyankameel86@gmail.com
Department of Information Technology, Delhi Technological University, India

**Abstract**: Social media is interactive, and interaction brings misinformation. With the growing amount of user-generated data, fake news on online platforms has become much frequent since the arrival of social networks. Now and then, an event occurs and becomes the topic of discussion, generating and propagating false information. Existing literature studying fake news primarily elaborates on fake news classification models. Approaches exploring fake news characteristics and ways to distinguish it from real news are minimal. Not many researches have focused on statistical testing and generating new factor discoveries. This study assumes fourteen hypotheses to identify factors exhibiting a relationship with fake news. We perform the experiments on two real-world COVID-19 datasets using qualitative and quantitative testing methods. This study concludes that sentiment polarity and gender can significantly identify fake news. Dependence on the presence of visual media is, however, inconclusive. Additionally, Twitter-specific factors like followers count, friends count, and retweet count significantly differ in fake and real news. Though, the contribution of status count and favorites count is disputed. This study identifies practical factors to be conjunctly utilized in the development of fake news detection algorithms.

# 1 Introduction

COVID-19 spread worldwide even faster than a human brain could imagine. Humans hardly even heard about it than before it turned to be the most fatal. After facing the catastrophic results only, many people became aware of it and started to ponder it. Talks about COVID-19 were everywhere and on everybody's minds and lips. Interactions about the hot topic have overwhelmed social networking platforms. Social media has now established its feet to feed information to people in the easiest way. The internet has been flooded with various types of information. But not everything that is on the internet is not reliable. Information that roams around on social media has not been validated and is merely people's ideas. Gradually these talks turned to be all sorts of fake news. With the feasibility of posting, sharing, and accessing the information on the web, its users can be quickly confounded with fake news. Fake news consists of every type of misinformation and disinformation. From the desks of politicians and public figures made the maiden attempt in spreading fake news worldwide, misleading people at large. It was the result of fake news that 5G towers in the UK turned into ruins. Fake news oozed out deadly political, social, religious, technological, environmental changes around the globe. It generated a sense of distrust among the people of the world. Enmity started grasping its enclosures. People claimed China to be the most causative element in spreading the disease. Detection of all sorts of talks that tend to be getting converted as fake news was the greatest need to lead the world into another mass destruction-like situation.

Fake news about the pandemic sprawled amongst various dimensions of society. One of these is the claiming of the remedial part of COVID-19. Enormous remedial approaches and suggestions started their part to play in contributing to fake news. "A pinch of turmeric or a drop of garlic juice could cure the fatal" was amongst the most prevailing unauthentic fake remedies. Poor perceptions, unproven methods, illogical claims, false figures, and alarming news overwhelmed the global information scenario. Social media platforms are well known for the spread of misinformation and denial of scientific literature [1]. False social media posts have also tricked users into relying on harmful and poisonous substances like weed, cannabis, and ethanol intake [2]. The rapid evolution of the COVID-19 pandemic has not permitted immediate and specific scientific data [3]. COVID-19 is not the only fake news generating event. In the past, there have been many instances that led to colossal misinformation spread on online social networks, such as the 2016 US presidential elections, Pizzagate, hurricane Harvey, etc. [4]. COVID-19, whereas, is one major event generating misinformation on a scale larger than any other events. This led the World Health Organization into coining the term "Infodemic," referring to the mass propagation of false news revolving around the pandemic.

Previous research has contributed variously to solving the fake news problem. Researchers from behavioral sciences have covered the factors involved in sharing and accepting fake news [5, 6, 7]. Others have investigated several factors like user demographics and background information [8]. Many studies have developed fake news detection algorithms [9, 10]. Such algorithms widely utilize news content, such as linguistic features, visual features, and network features. However, there is an absence of ideal classifiers, and most of the fake news characteristics are unidentified. In this paper, we identify several key factors associated with fake and real news on Twitter. We formulate fourteen hypotheses on the key elements and their direct and mediating relationship with fake news. These hypotheses are evaluated on two real-world datasets which contain tweets about the COVID-19 pandemic. MediaEval 2020 [11] is a benchmark dataset containing tweets pertaining to coronavirus and 5G conspiracy. CovidHeRA [12] is a collection of tweets associated with spreading health-related misinformation amidst the pandemic. The contribution of this paper is the analysis of characteristics that differentiate between fake and real news. We identify the following key factors: sentiment polarity, gender, media usage, follower count, friends count, status count, retweet count, and favorites count. Interdependence of factors like sentiment polarity, gender, and media usage are studied intensely. The relationship between fake news and these factors has not been studied in past research. We also extend the work of Parikh et al. [13] by demonstrating the relationship between fake news and particular sentiment polarities. This paper comes up with exciting outcomes suggesting important features demonstrating fake news dependence. The research bridges existing gaps in the literature and forms the basis for a new direction in fake news analysis. Our hypotheses shall be helpful in developing efficient fake news detection algorithms covering a wide range of fake news components.

The organization of this paper is as follows: Section 2 studies the existing literature in fake news and COVID-19 Infodemic. The survey provides insights into existing hypotheses and conclusions drawn upon fake news. Section 3 presents the research methodology explaining the datasets used and characteristics assumed for this study. Section 4 covers the results obtained by performing statistical tests on the datasets. Section 5 describes the insights drawn from the results and summarizes the acceptance/rejection of formulated hypotheses. Section 6 concludes the paper by discussing future directions.

## 2  Literature Review

This section discusses the progress in fake news and hypothesis domain presented by fellow researchers so far.

**Fake News:** The menace of fake news has been a challenging problem for information consumers. It has constantly been a topic of concern in the research society. Various studies have discussed the identification and detection of fake news on online social networks [14]. Past studies have focused on a vast dimension of fake news ranging between its origin, propagation, consumption, and impact [15]. In the recent era, various solutions have been proposed to detect fake news by the help of exploiting its textual [16], visual [17], and nodal features [18]. In contrast, studies pertaining to hypothesis formulation and testing are very few. There is limited literature available discussing the latest trends in online social networks highlighting vulnerabilities in fake news propagation and consumption. It is essential to formulate and discover dependent dimensions of fake news. Some studies have proposed important insights beneficial for fake news detection. For instance, Parikh et al. proposed hypotheses discussing the origin, proliferation, and tone of fake information [13]. They concluded that such misleading information is published more on lesser-known websites than the popular ones. In terms of proliferation or sharing, unverified users are more often shared on social media than by verified accounts. They also demonstrated that fake news has a specific tone or sentiment (positive, negative, or neutral) but did not conclude which type of particular tone is fake news mostly related to. Their study provides ways to form additional hypotheses, which is also a motivation for our work. Demographics and culture form the basis of theories proposed by Rampersad and Althiyabi [8]. They identified the established relationships between age and acceptance of fake news. It was noted that other demographics like gender and education played a more minor role in fake news acceptance. Another notable hypothesis confirmed that educated people are less likely to accept fake news. It was also observed that culture indirectly impacts the acceptance of fake news significantly. Works have highlighted the connection between Third Person Effect (TPE) and fake news sharing [19, 20]. Brewer et al. have drawn several conclusions towards readers' reactions to consuming fake news [21]. Horne et al. have distinguished between real and fake news based on stylistic and physiological features of the text [22]. In another work by Silverman and SingerVine, it was identified that 75% of the US adults accepted fake news as true [23]. Similarly, Bovet and Makse studied the fake news propagation on Twitter during the 2016 US presidential elections and explored its influence [24]. Altay et al. hypothesized the relation between users' reputation and fake news sharing [5]. They studied that very few people were indulged in sharing fake news and identified the causes of such behavior. They arrived at the conclusion that sharing fake news harmed people's reputations and resulted in trust issues, which is a significant reason for very few people being indulged in sharing fake news. Osatuyi and Hughes figured that the amount of information available on fake news platforms is lesser than real news [25]. Exploring the role of comments in identifying and rejecting fake news shows that users are less likely to accept fake news if they come across critical comments about the content [26].

**Infodemic:** With the outbreak of the COVID-19 pandemic, social media communication and interactions rose at a level greater than before. Global concerns about the disease brought the world together to share information on online social networks. Such large-scale propagation gave rise to a phenomenon- "Infodemic." In an early response, researchers approached this problem by analyzing various concerns and suggesting solutions to the issue. Moscadelli et al. [27] have investigated the topics about the pandemic most polluted with fake news. Calvillo et al. [28] have analyzed political associations with the discerning of fake news. Hypotheses linking the fake news belief structure to its acceptance, Kim and Kim [29] proposed that factors like source credibility, quality of information, receiver's ability, perceived benefit, trust, and knowledge decrease people's belief in fake news. Contrastingly, heuristic information, perceived risk, and stigma strengthen the confidence in fake news. Greene and Murphy [30] have discussed the likeliness of people sharing true or false stories on social media, establishing the association with their knowledge concerns. Another study that links conscience and ideology with infodemic sharing behavior is provided by Lawson and Kakkar [31]. Montesi [32] spreads light on the nature of infodemic and suggests that the harm caused by fake news is not health-related but more of a moral sort. Society, politics, and society are identified as the dominant infodemic themes. Building constructs over the Third Person Effect (TPE), Lui and Huang [33] have facts regarding the susceptibility and perception of fake news in the pandemic era. Similarly, Laato et al. [34] discuss the factors such as information sharing, information overload, and cyberchondria aiding fake news propagation. Experimenting on a Nigerian sample, Sulaiman [35] proposed no relationship between information evaluation and fake news sharing. With many hypotheses, Alvi and Saraswat [36] explored connections amongst various heuristic and systematic factors such as Sharing Motivation, Social Media Fatigue, Feel Good Factor. Fear Of Missing Out. News Characteristics, Extraversion, Conscientiousness, Agreeableness, Neuroticism, Trust, and Openness. As observed from the existing literature, past studies revolve around identifying psychological and behavioral factors that demonstrate any relationship with fake news. There is a research gap in characterizing features that could aid in distinguishing false information from real and serve as contributing factors to build fake news detection algorithms.

## 3  Research Methodology
### 3.1 Data

This study uses two publicly available benchmark datasets, MediaEval 2020 [11] and CovidHeRA [12]. MediaEval 2020 issued a benchmark dataset for its fake news detection task. The dataset consists of 5842 tweets classified into three classes: 5G coronavirus conspiracy, other conspiracy, and non-conspiracy. The tweets contain real and false information revolving around the COVID-19 pandemic. For this study, we classify these tweets into two coarse classes, with non-conspiracy tweets as real and the remaining tweets as fake. CovidHeRA is another benchmark dataset containing false tweets related to coronavirus and health. These tweets are a collection of fake remedies, preventive measures, treatments, and other health-related information spread across Twitter amidst the pandemic. Originally, the datasets consisted of tweet ids. To procure various characteristics of the tweets, the python library Tweepy is utilized. This scraping results in providing various information of the tweet and user content. This contextual information forms the basis of this study. To obtain the gender information of Twitter users, a gender predictor algorithm by Sap et al. [37] is used. Sentiments on the dataset are extracted using Microsoft's Text Analytics service. Sentiment scores are returned as values in the range of 0.0 to 0.1. A score between 0.0 to 0.3 signifies negative, 0.3 to 0.7 represents neutral and 0.7 to 1.0 represents positive sentiment. For media usage, we utilize the 'extended_entities' column from the scraped datasets. Sizes of both the datasets pertaining to each category are provided in tables 1, 2, and 3.

Table 1: Count of fake and real items with gender as a category

| Label | CovidHeRA | | | Mediaeval | | |
|---|---|---|---|---|---|---|
| | Male | Female | Total | Male | Female | Total |
| Fake | 1532 | 772 | 2304 | 929 | 837 | 1766 |
| Real | 42683 | 40104 | 82787 | 2011 | 2065 | 4076 |
| Total | 44215 | 40876 | 85091 | 2940 | 2902 | 5842 |

Table 2: Count of fake and real items with sentiment polarity as a category

|  | CovidHeRA | | | | Mediaeval | | | |
|---|---|---|---|---|---|---|---|---|
| Label | Negative | Neutral | Positive | Total | Negative | Neutral | Positive | Total |
| Fake | 1292 | 391 | 621 | 2304 | 1042 | 346 | 378 | 1766 |
| Real | 31004 | 24638 | 27145 | 82787 | 2320 | 690 | 1066 | 4076 |
| Total | 32296 | 25029 | 27766 | 85091 | 3362 | 1036 | 1444 | 5842 |

Table 3: Count of fake and real items with media usage as a category

|  | CovidHeRA | | | Mediaeval | | |
|---|---|---|---|---|---|---|
| Label | With | W/o Media | Total | With | W/o Media | Total |
| Fake | 150 | 2154 | 2304 | 289 | 1477 | 1766 |
| Real | 17700 | 65087 | 82787 | 791 | 3285 | 4076 |
| Total | 17850 | 67241 | 85091 | 1080 | 4762 | 5842 |

## 3.2 Research Hypotheses

To identify characteristics that distinguish fake news and real news and consequently identify fake news based on these characteristics, we have formed fourteen hypotheses based on the qualitative and quantitative variables present in the dataset. Past research to determine factors related to fake news is limited. To identify the dependence of social media misinformation, we identify and analyze eight key elements: sentiment polarity, gender, media usage, follower count, friends count, status count, retweet count, and favorite count. We assume that fake news characterization, propagation, and acceptance have a relationship with these factors, which can be consequently utilized in fake news detection. For a better understanding, each tweet labeled as fake/real in the datasets has specific characteristics mentioned above. It is crucial to examine which feasible aspects demonstrate a relationship with false tweets. We also aim to study if there are any significantly different factors between real and fake tweets. By establishing such relationships, we tend to describe certain features useful for real and fake tweet classification. As evident from the existing literature, very few features have been exploited by fake news detection algorithms. Now examining the stated features, we propose to add more of such contributing characteristics. Qualitative hypotheses $H_A$, $H_B$, and $H_C$, are tested to scrutinize the direct relationships between sentiment, gender, and media usage with fake news, respectively. Further, it is vital to analyze if the bias of one independent variable influences the bias of another independent variable. For example, to test whether or not it is the higher proportion of one categorical variable contributing to the higher proportion of another categorical variable. To do so, we construct six more qualitative hypotheses, $H_D$, $H_E$, $H_F$, $H_G$, $H_H$, and $H_I$. These nine hypotheses are tested using the Chi-square test of independence. The relationship is demonstrated in figure 1. To study quantitative variables, we formulate hypotheses $H_J$ to $H_N$ and perform Analysis of Means on each one of, also and calculate intervals. Figure 2 demonstrates the quantitative relationships.

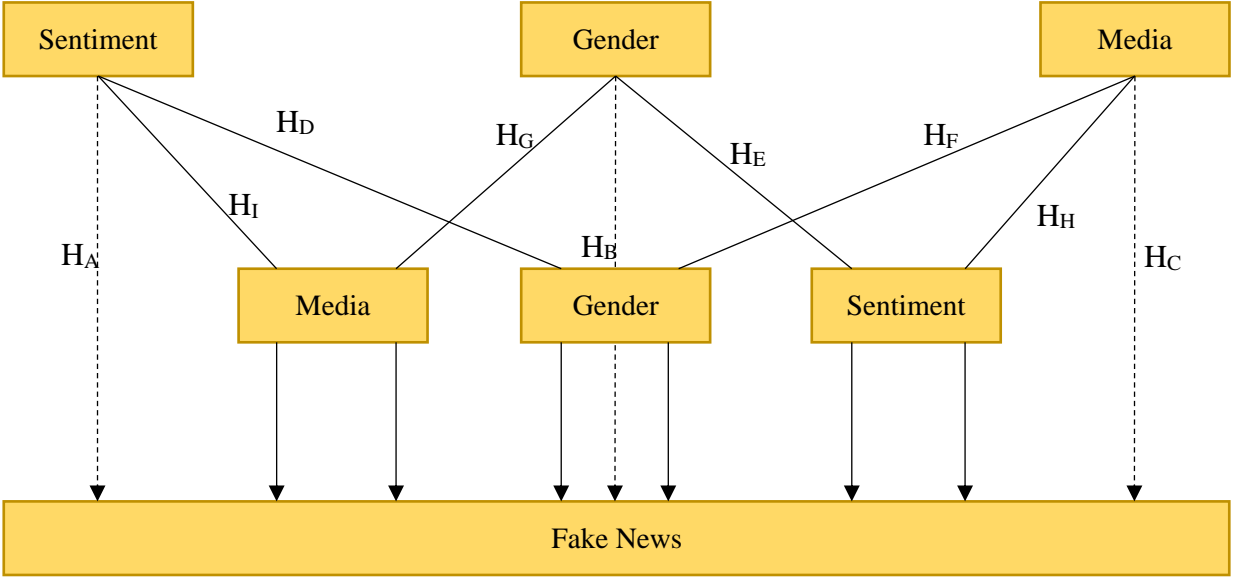

Figure 1: Factors determining fake news (qualitative hypotheses)

### 3.2.1 Qualitative Hypotheses and Factors

**Sentiment:** According to Parikh et al., it is widely assumed that most of the news spreading online is negative in terms of its linguistic tone. However, it has not been proven that fake news has a higher negative polarity than neutral or positive polarities. Parikh et al. noted that it was inconclusive to say if fake news had a bias towards a particular polarity. Following their assumption, $H_A$ forms the primary hypothesis to test if fake news has a tendency towards a specific sentiment polarity.

**$H_{A0}$: There is no bias in the proportion of different sentiments between fake news and real news.**
**$H_{A1}$: There is a significant bias in the proportion of different sentiments between fake news and real news.**

**Gender:** Rampersad and Althiyabi, examining a sample of Saudi Arabia, observed that gender has a weakly positive effect on the acceptance of fake news by people. The sample is specific to a particular demographic region. In our research, datasets consist of tweets from Twitter users across the globe. This helps to examine the assumptions on a universal scale. We test this hypothesis by using HB's statement to verify if there is a significant relationship between gender and false information.

**$H_{B0}$: There is no bias of the gender of users involved in fake news with respect to real news.**
**$H_{B1}$: There is a significant bias of gender of users involved in fake news with respect to real news.**

**Media:** Several fake news detection algorithms have been designed that detect whether a visual media in a piece of fake information is credible or not. We, hereby, analyze whether it can be stated solely based on the presence of visual media that a post/message is false. We categorize the datasets into two modalities: without and with visual media (pictures/videos). We try to analyze what data modality of social media posts contribute more/demonstrate bias towards misinformation using the statement $H_C$.

**$H_{C0}$: There is no bias of media usage in fake news with respect to real news.**
**$H_{C1}$: There is a significant bias of media usage in fake news with respect to real news.**

Based on the above three univariate hypotheses, we decide the mediating relationships among these factors and formulate multivariate hypotheses ($H_D$ to $H_I$) to determine whether bias in one of the above proportions is due to bias in proportions of the other variable.

**$H_{D0}$: There is no influence of bias in the proportion of a particular gender of the user on the bias in the proportion of sentiments in fake news with respect to real news.**
**$H_{D1}$: There is significant influence of bias in the proportion of a particular gender of the user on the bias in the proportion of sentiments in fake news with respect to real news.**

**$H_{E0}$: There is no bias in the proportion of a particular sentiment used in fake news between different gender of users.**
**$H_{E1}$: There is a significant bias in the proportion of a particular sentiment used in fake news between different gender of users.**

**$H_{F0}$: There is no bias in inducing a particular sentiment with media usage in fake news.**
**$H_{F1}$: There is a significant bias in inducing a particular sentiment with media usage in fake news.**

**$H_{G0}$: There is no bias in the usage of media amongst different sentiments used in fake news.**
**$H_{G1}$: There is a significant bias in media usage amongst different sentiments used in fake news.**

**$H_{H0}$: There is no relationship between a particular gender and media usage in fake news.**
**$H_{H1}$: There is a significant relationship between a particular gender and media usage in fake news.**

**$H_{I0}$: There is no bias in and media usage in fake news between different gender of users.**
**$H_{I1}$: There is a significant bias in and usage of media in fake news between different gender of users.**

### 3.2.2 Quantitative Hypotheses and Factors

Using the data scraped from Twitter, we decided on testing our hypotheses on five key factors, which can be categorized into three user/profile-specific features, i.e., the number of followers, friends, and statuses and two post-specific features, i.e., retweets count and favorites count. In our novel approach, we assume that these factors can be utilized in identifying the credibility of tweets, or in other words, labeling of tweets. Moreover, we assume that these factors impose an effect on fake news sharing and acceptance.

Followers and friends count determine the extent of reachability of a particular post or message within the user's social network who created it. A retweet is an action of sharing a particular tweet on one's timeline, which is done mainly by the follower of the user who created it and is visible to other Twitter users who turn the followers of the user who retweeted it. Retweet count determines the propagation and acceptance behavior of a fake post by checking the social reach. It is similar to the action "Share" on other social networks. It spreads a particular post to the user's social network. The larger the retweet count, the more likely the people reading the post will believe that particular piece of information and further spread it across the web. Status count corresponds to the number of total posts/retweets a specific user has posted since the creation of his account. Favorites are user markings made on a post a user would like to save for the future.

We determine the relationship between these quantitative variables and the label of the post, i.e., the relationship between the number of retweets and favorites of the post and the followers, friends, and status of the user who posted it, and it being real or fake. Since the source of misinformation can range from a random regular user to a credible account such as commercial news channels, journalists, or celebrities, it becomes difficult to assume any specific range for the count of these quantitative variables. Hence, we test based on a characteristic whether there is a significantly distinguishable bias in the values attributed to the mean and a confidence interval around it for each of these variables. In other words, the probability with which a post or a piece of information under examination can be labeled as fake or real based on its values of the above-mentioned quantitative variables.

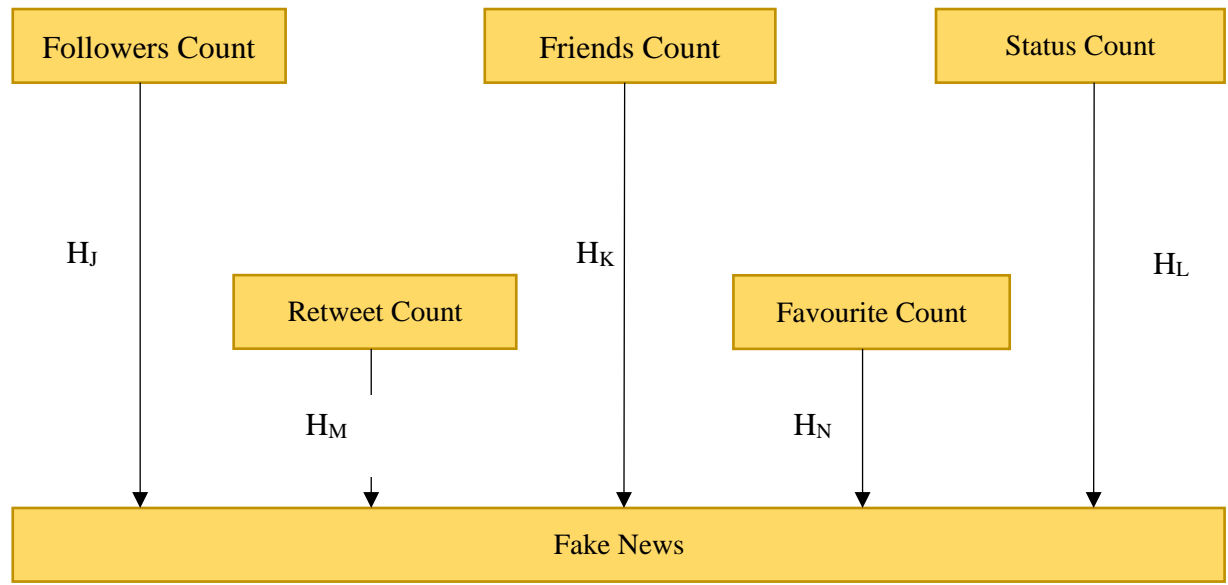

Figure 2: Factors determining fake news (quantitative hypotheses)

**$H_{J0}$: There is no bias of follower count in fake news.**
**$H_{J1}$: There is a significantly distinguishable bias of follower count in fake news.**

**$H_{K0}$: There is no bias of friends count in fake news.**
**$H_{K1}$: There is a significantly distinguishable bias of friends count in fake news.**

**$H_{L0}$: There is no bias of status count in fake news.**
**$H_{L1}$: There is a significantly distinguishable bias of status count in fake news.**

**$H_{M0}$: There is no bias of retweet count in fake news.**
**$H_{M1}$: There is a significantly distinguishable bias of retweet count in fake news.**

**$H_{N0}$: There is no bias of favorite count in fake news.**
**$H_{N1}$: There is a significantly distinguishable bias of favorite count in fake news.**

## 4  Results

To test on the nine hypotheses $H_A$ to $H_I$, which are formed upon the categorical variables, we use the Chi-Square test of independence alongside computing "Cramer's V," "Pearson's r," and "spearman's rho" values. Cramer's V value provides us with the strength of association between the nominal categorical variables for the conclusion arrived using the Chi-Square test. Its values range between 0 and 1. Pearson's r value signifies both the strength of association and the direction of the association between two continuous variables. Here direction indicates if one variable would increase or decrease with respect to change in another variable. Its values range from -1 to +1, where the value of -1 means that as one variable increases, the other decreases, and +1 means that as one variable increases, the other increases too. A value of 0 indicates no strength of association. Spearman's rho values differ from the outcomes of Pearson's r values by a feature that they can describe the correlation even when the variables do not have a linear association. It is also proof from the long tail of outlier values as it uses the ranks of the values of the variable. The values in the table 7 include degrees of freedom as df, Chi-Square test value as $\chi^2$, probability value as p-value and Cremer's V value, Pearson's r-value, and Spearman's rho. The first column in this table indicates the hypothesis to which the variables and their values belong to. From the first row of the same table, we observe that $\chi^2$ values for testing hypothesis $H_A$ with 2 degrees of freedom (df) for both CovidHeRA and MediaEval datasets are 352.963 and 17.103, respectively, and are more significant than critical value $\chi^2_c = 5.991$ with $p < 0.001$ (Significance level $\alpha = 0.05 = p_c$, critical p-value). This implies that there is a significant difference in proportions of sentiments used between Fake and Real news. But despite there being a substantial difference in ratios, low values of Cramer's V (less than 0.2), Pearson's r (between -0.20 and +0.20), and Spearman's rho (between -0.20 and +0.20) indicate weak association of label (news being fake or real) and the sentiment (sentiment being negative or neutral or positive). These values (Cramer's V, Person's r, and Spearman's rho) are low for all the hypotheses tested. Therefore, we rely on comparing Actual values from tables 1,2 and 3 with Expected values in tables 4,5 and 6, respectively, to determine the association between an independent and a categorical dependent variable, or in other words, the bias of fake news towards a specific or a group of categorical variables. On comparing table 2 and table 5, we observe that in both CovidHeRA and MediaEval datasets, Fake news with Negative sentiment has a higher Actual count (1292, 1042) with respect to Expected count (874.5, 1016.3) and Fake news with Positive sentiment has less Actual count (621, 378) with respect to Expected count (751.8, 436.5). Count of Neutral sentiment varies inversely in both datasets, with CovidHeRA showing reduced count and MediaEval showing an Increase. Similarly, we observe from the same tables that the Actual count of Real news with Negative sentiment is less than that of the Expected count in both datasets. The Actual count of Real news with Positive sentiment is greater than that of the Expected count in both the datasets. Therefore, we reject the Null hypothesis ($H_0$) of $H_A$ and observe that Fake news propagation during CoVID-19 has had a proportional bias towards Negative sentiment.

Table 4: Expected count of fake and real items with gender as a category

| Label | CovidHeRA | | | Mediaeval | | |
|---|---|---|---|---|---|---|
| | Male | Female | Total | Male | Female | Total |
| Fake | 1197 | 1107 | 2304 | 888.7 | 877.3 | 1766 |
| Real | 43018 | 39769 | 82787 | 2051.3 | 2024.7 | 4076 |
| Total | 44215 | 40876 | 85091 | 2940 | 2902 | 5842 |

Table 5: Expected count of fake and real items with sentiment polarity as a category

| Label | CovidHeRA | | | | Mediaeval | | | |
|---|---|---|---|---|---|---|---|---|
| | Negative | Neutral | Positive | Total | Negative | Neutral | Positive | Total |
| Fake | 874.5 | 677.7 | 751.8 | 2304 | 1016.3 | 313.2 | 436.5 | 1766 |
| Real | 31421.5 | 24351.3 | 27014.2 | 82787 | 2345.7 | 722.8 | 1007.5 | 4076 |

| | | | | | | | |
|---|---|---|---|---|---|---|---|
| Total | 32296 | 25029 | 27766 | 85091 | 3362 | 1036 | 1444 | 5842 |

Table 6: Expected count of fake and real items with media usage as a category

| | CovidHeRA | | | | Mediaeval | | |
|---|---|---|---|---|---|---|---|
| Label | With | W/o Media | Total | | With | W/o Media | Total |
| Fake | 483 | 1821 | 2304 | | 326.5 | 1439.5 | 1766 |
| Real | 17367 | 65420 | 82787 | | 753.5 | 3322.5 | 4076 |
| Total | 17850 | 67241 | 85091 | | 1080 | 4762 | 5842 |

From the second row of the table 7, we observe that $\chi^2$ values for testing hypothesis $H_B$ for both CovidHeRA and MediaEval datasets are 200.321 and 5.261, respectively, and are greater than critical value $\chi^2_c = 3.841$ with $p < 0.001$ and $p = 0.022$, respectively, both less than $\alpha = 0.05$. This implies a significant difference in proportions of the gender of users between Fake and Real news. By comparing Actual values with Expected values from table 1 and table 4 respectively, we observe that the Male gender has a greater Actual proportion in Fake news than the Expected proportion, and the Female gender has a higher Actual proportion involved in Real news than Expected Proportion, in both datasets. Therefore, we reject the Null hypothesis ($H_0$) for $H_B$ and observe a significant bias in the gender of users involved in CoVID-19 Fake news propagation.

To test for Hypothesis $H_C$, from third row of table 7, we observe that $\chi^2$ values for both CovidHeRA and MediaEval datasets are 298.995 and 7.765, respectively, and are greater than critical value $\chi^2_c = 3.841$ with $p < 0.001$ and $p = 0.006$ respectively, both less than $\alpha = 0.05$. For both the datasets, comparing the values of Actual and Expected Media usage from table 3 and table 6 respectively shows that Actual values for Fake news with media used is less than Expected values and the same is more in the case of Real news. Therefore, there is a significant difference in the proportion of Fake news and Real news propagation with media usage than the expected proportion, which leads us to reject the Null Hypothesis ($H_0$) for $H_C$.

The test for hypothesis $H_D$ involves comparing values from row four and row five of the table 7. From row 4, the $\chi^2$ values of Male gender from datasets CovidHeRA and MediaEval are 217.67 and 13.342, respectively, both higher than $\chi^2_c = 5.991$ and p values being $p < 0.001$ and $p = 0.001$, respectively, both less than $\alpha = 0.05$. From row 5, the $\chi^2$ value for Female gender from CovidHeRA dataset is 169.979, greater than the critical value $\chi^2_c = 5.991$ and the value of $p < 0.001$ is less than $\alpha = 0.05$. But for the same gender in the MediaEval dataset, the $\chi^2$ value turns out to be 5.503, which is less than $\chi^2_c = 5.991$, and the p-value of $p = 0.064 > \alpha = 0.05$ suggests contradictory inference from these two datasets. But since the MediaEval dataset gave both the $\chi^2$ and p values close to their respective critical values for female gender, we reject Null Hypothesis ($H_0$) for $H_D$ and conclude that there is a significant bias in proportion of sentiments used by users of both the gender and the bias in proportion of the user gender has no influence on the bias of proportion of sentiments.

Further, to identify towards which sentiment is the bias more by the users of both genders, we use results from rows six, seven, and eight of table 7 for testing hypothesis $H_E$. For the CovidHeRA dataset, the three rows mentioned above have $\chi^2$ value of 78.005, 13.65, and 146.509 for negative, neutral, and positive sentiment, respectively, which are all greater than $\chi^2_c = 3.841$ and their respective p values being $p < 0.001$ for all three, is less than $\alpha = 0.05$. Results from this dataset do not indicate the specific sentiment towards which the bias is more. However, we can infer that there is a significant difference in the proportion of each sentiment when compared to real news. Observing results from these three rows for the MediaEval dataset, we obtain $\chi^2$ value of 1.702, 4.82, and 0.411, for negative, neutral, and positive sentiments, respectively, where $\chi^2$ values for Negative and Positive sentiments are both less than $\chi^2_c = 3.841$ and for Neutral sentiment, the $\chi^2$ value is higher than $\chi^2_c$. The p values for these corresponding $\chi^2$ values are p = 0.192, p = 0.028 and p = 0.521, respectively. This shows no significant bias of the user gender on Negative and Positive sentiment as p values (0.192 and 0.521) obtained are greater than $\alpha = 0.05$. But for Neutral sentiment, we observe a bias as the p-value of 0.028 is less than $\alpha = 0.05$. Therefore, we reject the Null hypothesis ($H_0$) for $H_E$ and conclude that Fake news is more biased towards being sentiment Neutral, followed by being sentiment Negative, and show no significant difference in proportions of Real news towards being sentiment Positive.

For testing Hypothesis, $H_F$, the bias of usage of media to induce a particular sentiment in the propagation of COVID-19 Fake news, from table 7, the values from rows nine, ten and eleven for CovidHeRA dataset indicate $\chi^2$ values of 97.382, 59.615, and 124.61 for Negative, Neutral and Positive sentiment, respectively, with all of them being greater than $\chi^2_c = 5.991$ and with a p-value for each of them being p < 0.001, less than $\alpha = 0.05$ indicate rejection of Null Hypothesis ($H_0$) for $H_F$. For the MediaEval dataset, however, the $\chi^2$ values of 0.235, 2.399, and 20.077 for Negative, Neutral and Positive sentiments, respectively, with the former two being less than $\chi^2_c = 3.841$ and the latter being more excellent, and their respective p values being p = 0.628, p = 0.121 and p < 0.001 indicate that only for Positive sentiment, there is a significant difference of proportion in the usage of media for Fake news with respect to real news. From the contradictory results from the two datasets for Negative and Neutral sentiments, we understand that there is a bias produced by usage of media for only positive sentiment. Hence, we reject the Null Hypothesis ($H_0$) for $H_F$.

Table 7: Chi-square test on qualitative hypotheses

| (H) | variables ↓ | Datasets → CovidHeRA | | | | | | MediaEval | | | | | |
|---|---|---|---|---|---|---|---|---|---|---|---|---|---|
| | | df | $\chi^2$ | p value | $V^2$ | r | rho | df | $\chi^2$ | p value | $V^2$ | r | rho |
| $H_A$ | Label vs Sentiment | 2 | 352.963 | p < 0.001 | 0.004 | 0.047 | 0.048 | 2 | 17.10 | p < 0.001 | 0.003 | 0.037 | 0.032 |
| $H_B$ | Label vs Gender | 1 | 200.321 | p < 0.001 | 0.002 | 0.048 | 0.048 | 1 | 5.261 | p = 0.022 | 0.001 | 0.03 | 0.03 |
| $H_C$ | Label vs Media usage | 1 | 298.995 | p < 0.001 | 0.003 | 0.059 | 0.059 | 1 | 7.565 | p = 0.006 | 0.001 | 0.035 | 0.035 |
| $H_D$ | Label vs Sentiment (Gender - male) | 2 | 217.67 | p < 0.001 | 0.004 | 0.039 | 0.041 | 2 | 13.342 | p = 0.001 | 0.004 | 0.035 | 0.027 |
| $H_D$ | Label vs Sentiment (Gender - female) | 2 | 169.979 | p < 0.001 | 0.004 | 0.058 | 0.058 | 2 | 5.503 | p = 0.064 | 0.001 | 0.038 | 0.035 |
| $H_E$ | Label vs Gender (Sentiment - Negative) | 1 | 78.005 | p < 0.001 | 0.002 | 0.049 | 0.049 | 1 | 1.702 | p = 0.192 | 0.001 | 0.023 | 0.023 |
| $H_E$ | Label vs Gender (Sentiment - Neutral) | 1 | 13.65 | p < 0.001 | 0.001 | 0.023 | 0.023 | 1 | 4.82 | p = 0.028 | 0.004 | 0.068 | 0.068 |
| $H_E$ | Label vs Gender (Sentiment - Positive) | 1 | 146.509 | p < 0.001 | 0.005 | 0.072 | 0.072 | 1 | 0.411 | p = 0.521 | 0 | 0.016 | 0.016 |
| $H_F$ | Label vs Media usage (Sentiment - Negative) | 1 | 97.382 | p < 0.001 | 0.003 | 0.055 | 0.055 | 1 | 0.235 | p = 0.628 | 0 | -0.008 | -0.008 |
| $H_F$ | Label vs Media usage (Sentiment - Neutral) | 1 | 59.615 | p < 0.001 | 0.002 | 0.048 | 0.048 | 1 | 2.399 | p = 0.121 | 0.002 | 0.048 | 0.048 |
| $H_F$ | Label vs Media usage (Sentiment - Positive) | 1 | 124.61 | p < 0.001 | 0.004 | 0.066 | 0.066 | 1 | 20.077 | p < 0.001 | 0.013 | 0.117 | 0.117 |
| $H_G$ | Label vs Sentiment (Media not used) | 2 | 267.346 | p < 0.001 | 0.003 | 0.042 | 0.044 | 2 | 7.223 | p = 0.027 | 0.001 | 0.01 | 0.005 |
| $H_G$ | Label vs Sentiment (Media used) | 2 | 61.585 | p < 0.001 | 0.003 | 0.045 | 0.042 | 2 | 25.154 | p < 0.001 | 0.023 | 0.145 | 0.141 |
| $H_H$ | Label vs Gender (Media not used) | 1 | 193.333 | p < 0.001 | 0.003 | 0.053 | 0.053 | 1 | 5.561 | p = 0.018 | 0.001 | 0.034 | 0.034 |
| $H_H$ | Label vs Gender (media used) | 1 | 5.472 | p = 0.019 | 0 | 0.017 | 0.017 | 1 | 0.345 | p = 0.557 | 0 | 0.017 | 0.017 |
| $H_I$ | Label vs Media usage (Gender - male) | 1 | 209.649 | p < 0.001 | 0.005 | 0.069 | 0.069 | 1 | 5.664 | p = 0.017 | 0.001 | 0.043 | 0.043 |
| $H_I$ | Label vs Media usage (Gender - female) | 1 | 87.831 | p < 0.001 | 0.002 | 0.046 | 0.046 | 1 | 2.529 | p = 0.112 | 0.001 | 0.03 | 0.03 |

From rows twelve and thirteen of table 7, we test for Hypothesis $H_G$ to observe a bias of proportion of sentiment caused when media is used and when it is not used, respectively. For CovidHeRA dataset, for with usage of media (row 12) and without the usage of media row (13), $\chi^2$ values of 267.346 and 61.585, respectively, both less than $\chi^2_c = 5.991$ and their respective p values of $p < 0.001$ each for both being less than $\alpha = 0.05$, suggest that there is a difference in the proportion of sentiment used in Fake news with respect to Real news. Similar inference can be obtained from MediaEval dataset, in which, with the usage of media (row 12) and without the use of media row (13) have $\chi^2$ values of 7.223 and 25.15, respectively, both less than $\chi^2_c = 5.991$ and their respective p values of $p = 0.027$ and $p < 0.001$, both being less than $\alpha = 0.05$. Hence, there is a bias induced in the proportions of sentiment in Fake news with respect to Real news by usage and non-usage of media, and therefore we reject the Null Hypothesis ($H_0$) for $H_G$.

Further, from rows fourteen and fifteen of table 7, we test for Hypothesis $H_H$ to check for bias in proportion of gender of Fake news with respect to Real news is influenced by bias in usage of media. For CovidHeRA, we obtain $\chi^2$ values of 193.333 and 5.472 for "media used" and "media not used", respectively, both greater than $\chi^2_c = 3.84$ with their respective p values being $p < 0.001$ and $p = 0.019$, both less than $\alpha = 0.05$. For the MediaEval dataset, for the same rows, we obtain $\chi^2$ values of 5.561 and 0.345 and p values of $p = 0.018$ and $p = 0.557$ for "media used" and "media not used," respectively. We observe that for "media not used," the test shows the opposite result with that compared from CovidHeRA dataset, meaning that there is no difference in the proportion of user's gender when media is not used in Fake news propagation, with respect to Real news propagation. These contradictory results make Hypothesis $H_H$ inconclusive.

Table 8: Descriptive statistics of CovidHeRA(C) and MediaEval dataset(M)

| Statistics | Followers | | Friends | | Retweets | | Status | | Favorites | |
|---|---|---|---|---|---|---|---|---|---|---|
| | Fake | Real | Fake | Real | Fake | Real | Fake | Real | Fake | Real |
| Mean (C) | 5421.657 | 63656.21 | 3181.374 | 2293.652 | 154.132 | 628.718 | 56262.98 | 46189.4 | 2.238 | 7.766 |
| Standard Error (C) | 445.735 | 3847.024 | 149.150 | 35.658 | 38.207 | 17.616 | 2269.618 | 497.010 | 0.255 | 0.490 |
| Median (C) | 1742.5 | 21733 | 953 | 605 | 52 | 173 | 17180 | 9238 | 1 | 2 |
| Mode (C) | 706 | 7810 | 775 | 209 | 18 | 28 | 3145 | 1760 | 0 | 0 |
| Standard Deviation (C) | 21395.32 | 1106894 | 7159.233 | 10259.94 | 1833.94 | 5068.56 | 108941.7 | 143003.4 | 12.261 | 141.126 |
| Sample Variance (C) | 4.58E+08 | 1.23E+12 | 51254619 | 1.05E+08 | 3.36E+06 | 2.57E+07 | 1.19E+10 | 2.04E+10 | 150.344 | 19916.57 |
| Count (C) | 2304 | 82787 | 2304 | 82787 | 2304 | 82787 | 2304 | 82787 | 2304 | 82787 |
| Confidence Level(95.0%) (C) | 874.085 | 7540.138 | 292.483 | 69.890 | 74.886 | 34.527 | 4450.709 | 974.136 | 0.500 | 0.961 |
| Mean (M) | 23255.37 | 99511.34 | 3012.989 | 1999.394 | 260.701 | 644.781 | 38369.96 | 55846.16 | 679.669 | 2244.092 |
| Standard Error (M) | 9979.619 | 11302.57 | 302.646 | 159.182 | 71.410 | 61.498 | 1787.307 | 1744.372 | 201.229 | 224.780 |
| Median (M) | 4711.5 | 37160.5 | 733 | 609 | 60 | 155 | 12955 | 18305.5 | 48 | 292 |
| Mode (M) | 1180 | 12548 | 650 | 138 | 77 | 92 | 547 | 446 | 48 | 177 |
| Standard Deviation (M) | 419381.5 | 721596.4 | 12718.35 | 10162.77 | 3000.955 | 3926.28 | 75109.43 | 111366.9 | 8456.42 | 14350.77 |
| Sample Variance (M) | 1.76E+11 | 5.21E+11 | 1.62E+08 | 1.03E+08 | 9005729 | 15415676 | 5.64E+09 | 1.24E+10 | 71511039 | 2.06E+08 |
| Count (M) | 1766 | 4076 | 1766 | 4076 | 1766 | 4076 | 1766 | 4076 | 1766 | 4076 |
| Confidence Level(95.0%) (M) | 19560.05 | 22153.04 | 593.583 | 311.998 | 140.058 | 120.570 | 3505.461 | 3419.921 | 394.672 | 440.692 |

From the values in rows 16 and 17 in table 7, for the CovidHeRA dataset, both genders show that there is a difference in the proportion of media used for Fake news propagation with respect to Real news. This can be observed as the $\chi^2$ values of 209.649 and 87.831 for the male and female gender, respectively, are both greater than $\chi^2_c = 3.84$, and their respective p values, both $p < 0.001$ is more diminutive than $\alpha = 0.05$. In the MediaEval dataset, we observe from rows 16 and 17 of table 7 that while users of Male gender with $\chi^2$ value of 5.664 and $p = 0.017$ show difference in the proportion of media used for Fake news with respect to Real news, but for Female gender, indifference in proportions of usage of media in Fake news with respect to real news is observed as the $\chi^2$ value of 2.529 is less than $\chi^2_c = 3.84$ and its p-value of $p = 0.112$ is more remarkable than $\alpha = 0.05$. Therefore, for Hypothesis $H_I$, we cannot come to any conclusive decision.

For the Quantitative variables, we plot the data distribution around the mean with a 95% confidence interval. This will distinguish the central values of the variables and help us determine the strength of the distinguishment, i.e., the smaller the upper and lower bound distance from the mean, the more the reliance on these values representing the true mean value of the population. Table 8 provides descriptive statistics of both the datasets. From fig 3 and fig 4, for CovidHeRA and MediaEval dataset, respectively, we observe that users who propagated Fake news have a smaller number of followers than the users with Real news. The mean values for Fake news for these datasets are 5421.65 and 23255.37 in the order mentioned above. These are distinct from the mean number of followers of Real news 63656.21 and 99511.34 for the two datasets. We also observe that there is a significant bias in the number of followers of the users of Fake news and Real news as the range of 95% CI for mean do not overlap for Fake news and Real news and therefore attributing a label to a piece of information on Twitter by comparing, the number of followers of the user who shared it with the mean range of these plots can be done more accurately.

From the plots of the number of Friends in fig 3 and fig 4 for CovidHeRA and MediaEval datasets, respectively, the previously mentioned inference becomes much more robust as not only the 95% CI bounds remain distinct for Fake news and real news, but also the closer proximity of the value of mean for a particular label in both datasets shows the repeatability of the trend. The mean value for Fake news in CovidHeRA and MediaEval dataset is 3181.374 and 3012.989, respectively, and the same for Real news in these datasets is 2293.652 and 1999.394, respectively. There is a significant bias in the mean number of friends for users who propagated Fake news compared to the number of friends of users who propagated real news.

The plots from fig 4 and fig 4 for the number of retweets have similar mean value for fake news and real news. For Fake news, the mean values of 154.132 and 260.701 for the two datasets, and Real news, the mean values are 628.718, and 644.781 show the closeness within the label and distinction between the labels. Therefore, this bias can prove helpful to label a piece of information based on its proximity to one of the mean values 95% CI interval.

For the "number of statuses" variable, the 95% CI interval for mean and the mean value for Fake news and Real news alternate between the two datasets. Therefore, we cannot come to any specific conclusion using the information of this variable of a particular information sample despite there being bias in the mean values between the Labels. The same conclusion can be drawn for "number of favorites" as the ranges in both the datasets are significantly different amongst the same variable. Hence, any information about this variable in a sample information under test cannot be determined as Fake or Real.

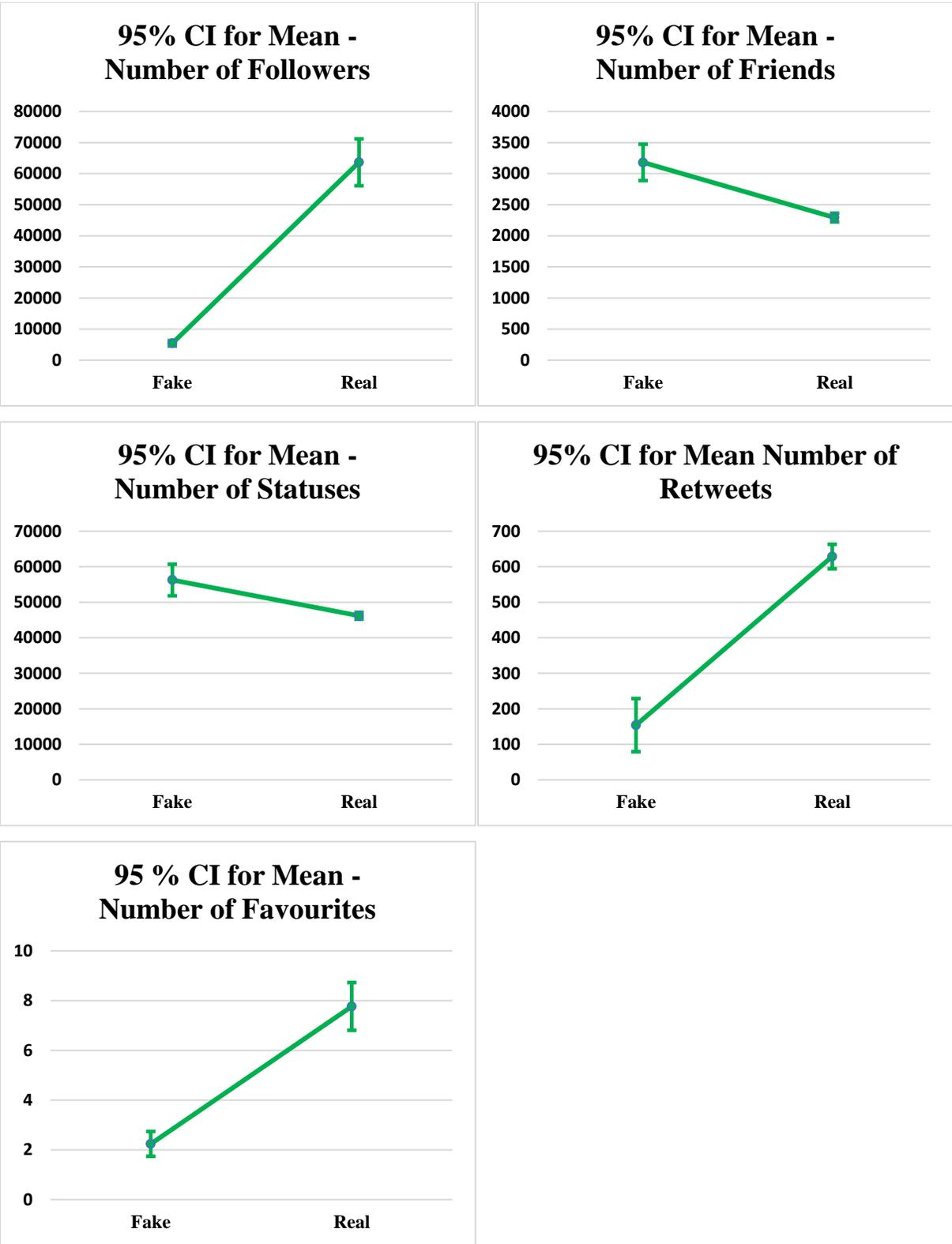

Figure 3: 95% Confidence Interval for quantitative factors on CovidHeRA dataset

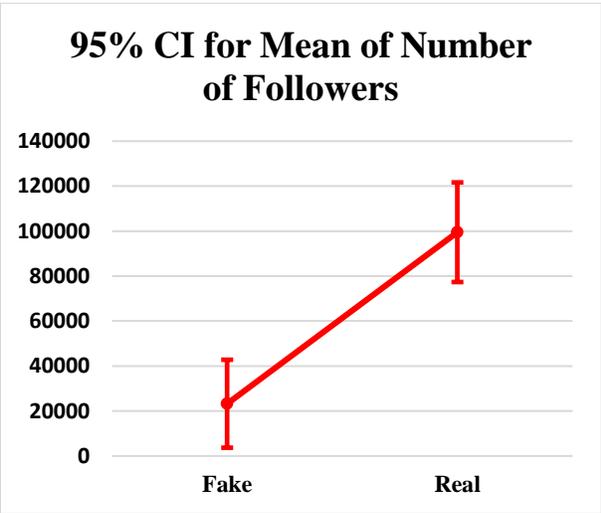
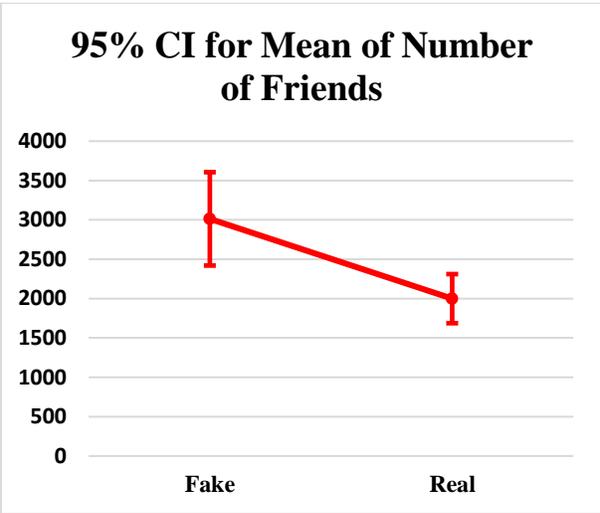
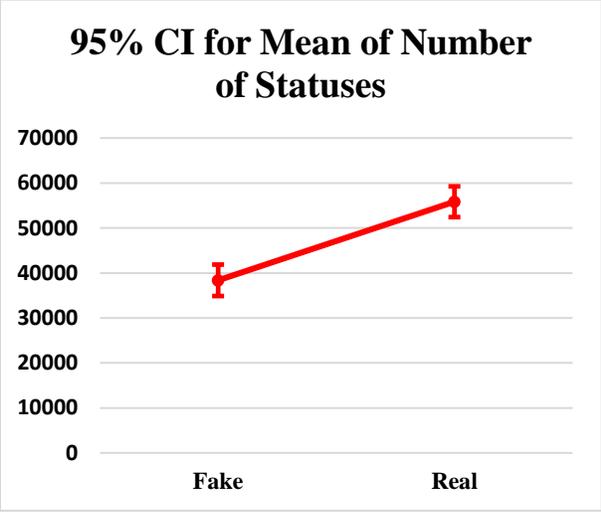
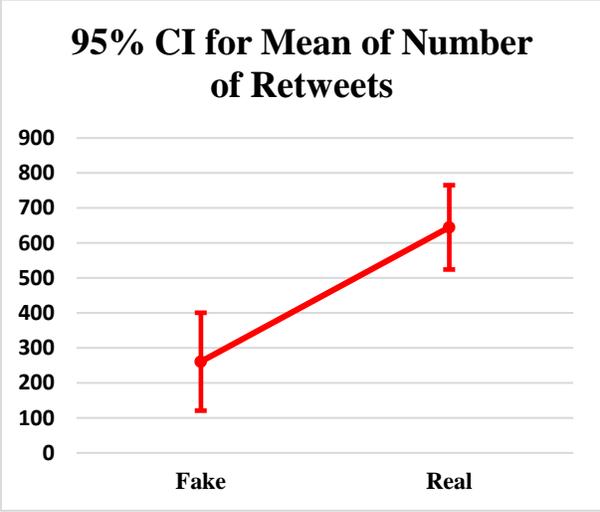
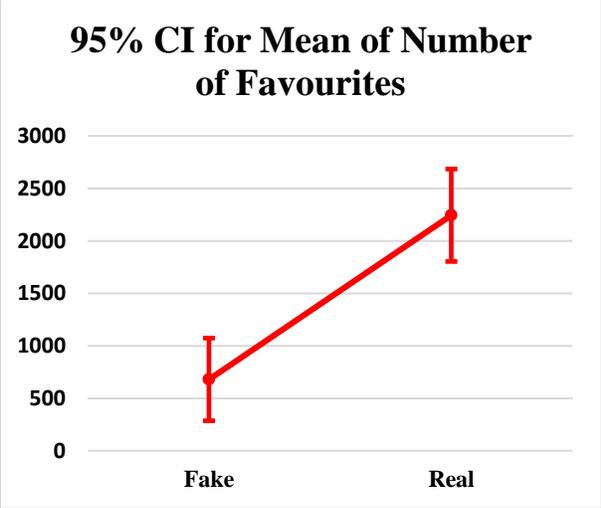

Figure 4: 95% Confidence Interval for quantitative factors on MediaEval dataset

## 5 Discussion

Fake news on social media is a menace hard to identify and characterize. It is unclear which factors are helpful in distinguishing between real and fake news. Past literature has identified several psychological and behavioral features associated with fake news propagation and acceptance. Little research has been done in identifying key factors characterizing fake news. This study delves deep into factor analysis and their interdependence. We examine how certain factors influence fake news detection and propagation on Twitter. Table 9 summarizes the results of all hypotheses considered.

Table 9: Summary Table

| Hypotheses | Results |
|---|---|
| $H_A$: Bias of sentiment in fake news with respect to real news. | Reject Null Hypothesis |
| $H_B$: Bias of the gender of users involved in fake news with respect to real | Reject Null Hypothesis |
| $H_C$: Bias of media usage in fake news with respect to real news. | Reject Null Hypothesis |
| $H_D$: Bias in the proportion of a particular gender of the user on the bias in the proportion of sentiments in fake news with respect to real news. | Reject Null Hypothesis |
| $H_E$: Bias in the proportion of a particular sentiment used in fake news between different gender of users. | Reject Null Hypothesis |
| $H_F$: Bias of inducing a particular sentiment with the usage of media in fake | Reject Null Hypothesis |
| $H_G$: Bias in the usage of media amongst different sentiments used in fake | Reject Null Hypothesis |
| $H_H$: Relationship between a particular gender and media usage in fake news. | Inconclusive |
| $H_I$: Bias in and usage of media in fake news between different gender of users. | Inconclusive |
| $H_J$: Significantly distinguishable bias of "follower" count in fake news. | Reject Null Hypothesis |
| $H_K$: Significantly distinguishable bias of "friends" count in fake news. | Reject Null Hypothesis |
| $H_L$: Significantly distinguishable bias of "status" count in fake news. | Fail to Reject Null Hypotheses |
| $H_M$: Significantly distinguishable bias of "retweet" count in fake news. | Reject Null Hypothesis |
| $H_N$: Significantly distinguishable bias of "favorite" count in fake news. | Fail to Reject Null Hypotheses |

In our qualitative hypotheses $H_A$, it is assumed that there is a bias in the proportions of sentiment (linguistic tone) in fake news. Although, the central polarity of bias was unclear. With our study on two COVID-19 specific datasets, we found a strong bias of fake news towards neutral sentiment followed by negative sentiment with respect to real news, which is proved by the results of our first hypothesis. In the second hypothesis, $H_B$, we tested the bias in the proportion of gender in fake news. The results predicted that there is a strong bias of the male gender towards fake news propagation with respect to real news. Now the influence of the gender ratio of Twitter users is not taken into account as the test is performed to distinguish characteristics of real news and fake news. Any sort of this influence is assumed to affect both types of news equally and nullify its effect. In other words, the speculated gender ratio of 6.85:3.15 should be observed in any random sample collection of tweets. Hence, we directly compare the actual ratio from the dataset without considering the deviation from the speculated ratio. In our datasets, the proportion of tweets (both real and fake) with media is more minor than tweets without media. From the chi-square test results on hypothesis $H_C$, we find that the proportion of fake news with media is significantly less than expected and substantially more than anticipated for real news with media. Further, we explore if the bias in proportions of one category amongst sentiment, gender, and media usage, is significantly influenced by the bias in proportions of these categories. From the test for Hypothesis $H_D$, we find that Fake news shared by both male and female gender show bias in proportion of sentiment. The result for hypothesis $H_E$ indicates that this bias is towards fake news being sentiment Neutral, followed by sentiment negative, with respect to real news. This supports our Hypothesis $H_A$. Further, from the results of testing Hypothesis $H_F$ and $H_G$, $H_G$ concludes that there is a bias of sentiment in both "with" and "without" media usage. From $H_F$, we conclude that this bias in fake news propagation is proportional to using positive sentiment. For the remaining combination of gender and media usage, from the results of hypotheses $H_H$ and $H_I$, it cannot be concluded if there is a mutual influence of Media usage and gender of the user in the bias observed in Hypothesis $H_B$ and $H_C$ due to the contradictory results from the two datasets. In Hypothesis $H_H$, the contradictory results for "media used" and for $H_I$, the contradictory results for "Female" gender.

From the quantitative variables, we observe a significant distinguishable difference in the mean number of followers, friends, and retweets for fake and real news. The smaller value of mean for followers can be attributed to why most Real news sources are official media channels and celebrity users who share information on Twitter. In contrast, fake news comes mostly from regular Twitter users who do not have such a huge following. Similar reasons can be attributed to a smaller mean value for retweets of fake news. For the larger value of mean for the number of friends, we understand that the users who propagate fake news are involved in more mutual social connections. Understandably, celebrities and official media sources, when compared to active regular Twitter users, do not have many mutual connections that Twitter classifies as "friends" and, therefore, the resulting smaller value of the mean. The confidence interval for mean for each of these plots acts as a range for true mean for fake and real news and can be used to identify any sample of data by comparing its mean to the 95% CI for the mean of these plots. The non-distinguishable mean value and reverse in the plotted trend for the number of statuses posted by the users who propagated fake and real news and the difference of range for the mean of the number of users who favorited the tweet between the two datasets make these variables unsuitable for classification of the label for the tweet.

## 6 Conclusion

Fake information on social platforms has constantly been increasing. In the state of the COVID-19 pandemic, this problem has grown at an exponential rate globally. The pandemic is one major event generating misinformation and promoting its consumption through social networks worldwide. In the absence of a holistic fake news detection model, it is unclear what factors can be used to identify misinformation. Very few past works are dedicated to identifying such factors. In this work, we examined several factors from two Twitter datasets, MediaEval 2020 and CovidHeRA, using fourteen hypotheses $H_A$ to $H_N$. The study uses Chi-square tests for nine qualitative theories ($H_A$ to $H_I$), whereas for five quantitative tests ($H_J$ to $H_N$), we have calculated Confidence Intervals using Analysis of Means. Observations from this study unravel specific characteristics to distinguish fake news from real news. These new findings pave the way for future research and development of fake news detection algorithms. We motivate fellow researchers to design algorithms that utilize the discovered dependencies using their combined decisions. Also, we encourage to discover more identifiers that can characterize false information present online ubiquitously. This study provides a new dimension to the existing literature in the fake news domain.

## Acknowledgments

We thank Mihir P Mehta (Indian Institute of Management Raipur) for his support and feedback in the experimental setup.